# Graphene Nanoribbons with Smooth Edges Behave as Quantum Wires

Xinran Wang,[1,2,†] Yijian Ouyang,[3] Liying Jiao,[1] Hailiang Wang,[1] Liming Xie[1], Justin Wu,[1] Jing Guo,[3] and Hongjie Dai[1,*]

[1] Department of Chemistry, Stanford University, Stanford, CA 94305, USA

[2] National Laboratory of Microstructures, School of Electronic Science and Engineering, Nanjing University, Nanjing 210093, China

[3] Department of Electrical and Computer Engineering, University of Florida, Gainesville, FL, 32611, USA

[†] Current address: Department of Material Science and Engineering, University of Illinois at Urbana-Champaign, Urbana, IL 61801, USA

[*] Correspondence to hdai@stanford.edu

Graphene nanoribbons with perfect edges are predicted to exhibit interesting electronic and spintronic properties[1-4], notably quantum-confined bandgaps and magnetic edge states. However, graphene nanoribbons produced by lithography have, to date, exhibited rough edges and low-temperature transport characteristics dominated by defects, mainly variable range hopping between localized states in a transport gap near the Dirac point[5-9]. Here, we report that one- and two-layer nanoribbons quantum dots made by unzipping carbon nanotubes[10] exhibit well-defined quantum transport phenomena, including Coulomb blockade, Kondo effect, clear excited states up to ~20meV, and inelastic co-tunnelling. Along with signatures of intrinsic quantum-

**confined bandgaps and high conductivities, our data indicate that the nanoribbons behave as clean quantum wires at low temperatures, and are not dominated by defects.**

In this study, we focused on $w$~10-20nm graphene nanoribbons with expected bandgaps in the range of $E_g \sim \dfrac{1eV}{w(nm)}$ ~50-100meV (Ref. 2). Recent transmission electron microscopy (TEM, Fig. 1c), aberration-corrected TEM (Fig. S1, Ref. 11) and scanning tunnelling microscopy (STM) measurements[12] revealed that an appreciable fraction of nanoribbons in our samples exhibited smooth edges with little roughness while some exhibited edge roughness on the order of ~1 nm. About 70% nanoribbons in our samples are non-AB-stacked bi-layer ribbons (Fig. S1), with ~10% single layer ribbons[11]. The electrical properties of a large number of nanoribbons measured exhibited variability, and a fraction of nanoribbons with short lengths (<100 nm) showed high conductance up to ~$7e^2/h$ (Fig. S5) and 'clean' quantum transport characteristics at low temperatures. Fig. 2a plots the room temperature conductance-backgate voltage $G$-$V_{gs}$ characteristics of a high quality nanoribbon device (called 'GNR1') with ribbon width $w$~14nm and channel length of $L$~86nm (Fig. 2a lower inset). The nanoribbon showed a topographic height of ~1.0nm after electrical annealing to remove physisorbed species[13], corresponding to either a single-layer or bi-layer nanoribbon[10,13]. The device exhibited a high p-channel conductance of $G > \dfrac{4e^2}{h}$ at room temperature (Fig. 2a). The resistance mainly came from quantum resistance at the contact of the graphene nanoribbon, and we estimated our contact transparency to be ~70% near the on-state. The conductivity $\sigma = G\dfrac{L}{w} \sim 0.97 mS$ and the calculated peak field-effect mobility $\mu = \dfrac{dG}{dV_{gs}}\dfrac{L^2}{C_g} \sim 1600 cm^2/Vs$ (the gate capacitance $C_g$=0.41aF was calculated by three-dimensional electrostatic simulation[14]) were much higher than those of previously reported nanoribbons ($\sigma \sim 0.1 - 0.3 mS$, $\mu \leq 700 cm^2/Vs$) with similar widths and number of layers ($\leq$ 2)[10,15-17]. Note that for short channel devices, the so called "ballistic mobility" and parasitic resistance could make the extracted mobility value lower than the mobility due to scattering[18].

The p-channel conductance of the nanoribbon increased as it was cooled from 290K to 50K (Fig. 2a upper inset, Ref. 20). At low temperatures ($< \sim 60K$), conductance at the Dirac point exhibited a drastic ($\sim 100$ fold) dip in a narrow gate range ($\Delta V_{gs} \sim 2V$) without any resonance-like sharp peaks due to localized states within the dip[5] (Fig. 2b, Fig. S6), suggesting an intrinsic bandgap of the nanoribbon[1,2] rather than the defect-induced transport gap (see Supplementary Information for our control experiments on lithographic ribbons)[5-9]. Considering the asymmetrical Schottky barriers for electrons and holes at the Pd contacts, we employed non-equilibrium Green's function (NEGF) approach to fit the experimental minimum conductance as a function of temperature to extract $E_g \sim 72 \pm 18$meV for this $w \sim 14$nm ribbon (Fig. S2, see Supplementary Information for details).

At a base temperature of 2K, the p-channel conductance of GNR1 was above $\frac{3e^2}{h}$ (Fig. 2b inset). The conductivity was up to $\sim 20$ times higher than previous nanoribbons with similar widths at low temperatures[5-9,16]. Near the bandgap, the nanoribbon behaved as a single quantum dot confined between the leads, and charge transport was through single electron charging[21]. We observed two prominent large diamonds (size $\sim 60$-70meV) near zero $V_{gs}$, presumably corresponding to the bandgap region[22], but the origin of two large center diamonds was unclear. A single large central diamond (with the size of $E_g$ plus charging energy) corresponding to the bandgap separating the electron and hole branches was expected, as in the case of semiconducting carbon nanotubes[22]. We note, however, that the appearance of two central diamonds varied in different cool-downs. In another cool down of the same ribbon, a single large diamond was observed (Fig. S3). Also, gate-switching events (Fig. 2c, S2) appeared common for nanoribbon devices due to sudden change in the charge environment of the nanoribbons. These observations suggested that less intrinsic factors could be at play possibly involving mobile adsorbates or charge impurities on or near the nanoribbons. Such effects have been suggested to induce mid-gap states in graphene nanoribbons[23], which could cause deviation from the single central diamond expected for the bandgap region.

On both sides of the bandgap region, regular Coulomb-blockade diamonds associated with charging through a single graphene nanoribbon quantum dot (suggested by closed periodic diamonds, in clear contrast to dots in series or parallel in previous nanoribbons[6,7]) were observed with the number of holes and electrons in the dot assignable (Fig. 2d, e). We noticed slight asymmetry for Coulomb diamonds in electron and hole branches, probably due to asymmetric tunnel barriers due to high work function Pd contact. The size of the diamonds along the bias voltage $V_{ds}$ axis in the electron branch (n-channel) was $E_{add} \sim 28$meV (Fig. 2d), consistent with the charging energy $E_c = \dfrac{e^2}{C} \sim 24.7 meV$, where $C = C_g + C_s + C_d \approx 6.48 aF$ was the total capacitance of the quantum dot[22,24] ($C_s$ and $C_d$ were source and drain capacitances respectively, and gate capacitance $C_g = \dfrac{e}{\Delta V_{gs}} = 0.43 aF$ based on the size of the diamonds $\Delta V_{gs} \sim 0.37$V along $V_g$, very close to that from the 3D electrostatic simulation).

We observed several discrete lines parallel to the edges of Coulomb diamonds attributed to transport through discrete excited states in the nanoribbon quantum dot[25,26] due to quantization along the length of the nanoribbons[22,27]. The first two measured energy levels outside diamond 1 could be assigned as the first and second excited state $\Delta\varepsilon_{21} = 3.6 meV$ and $\Delta\varepsilon_{31} = 16.2 meV$ respectively, where $\Delta\varepsilon_{n1} = \varepsilon\left(k_{\parallel}(n)\right) - \varepsilon\left(k_{\parallel}(1)\right)$ is the single-particle level above the ground state. In light of the uncertainties in the nanoribbon structures (number of layers and edge structures), we employed a simple model based on quantization of tight-binding Hamiltonians in the width and transport directions with particle-in-a-box boundary conditions to qualitatively understand the excited states energy (see Supplementary Information for details). As shown in Table S1, the calculated $E_g$ and $\Delta\varepsilon_{n1}$ for single- and several non-AB-stacked bi-layer nanoribbons were qualitatively in agreement (within a factor of ~3) with experiments. Quantitative comparison, however, was not possible due to the lack of detailed structural information.

In the hole branch, Kondo effect[28,32] was observed at 2K as enhanced conductance at zero bias inside the odd hole-number Coulomb diamonds (Fig. 2e, see the Kondo ridge or zero-bias horizontal lines in the 2D conductance plot). The differential conductance at zero bias showed the pairing of peaks, with non-zero intravalley conductance (Fig. 2e). The Kondo resonances were attributed to exchange interaction between a localized electron spin in the quantum dot and the delocalized electron spins in the metal leads. In the odd-number diamonds, the unpaired spin can form a spin singlet with electrons in the leads to give high conductance[28]. We can roughly estimate the Kondo temperature ($T_K$) by the bias at which the Kondo resonance is suppressed in the Kondo ridges[34]. From the $G$ vs. $V_{ds}$ plot in Fig. 2e inset, this energy scale is on the order of 1meV (the width of the Kondo resonance peak near zero bias is ~2mV), corresponding to $T_K$~10K, which is about an order of magnitude higher compared to carbon nanotube quantum dots[34,35]. Recently, $T_K$ was found as high as ~30-90K in defective graphene, attributed to strong coupling of Dirac electrons to magnetic defects[36].

Several other interesting transport phenomena were also present in GNR1. In Fig. 2e, there were finite (non-zero) conductance regions inside several even hole-number diamonds beyond horizontal lines intersecting the excited state lines at the edge of the Coulomb diamonds. These were attributed to inelastic co-tunneling of carriers through an excited state when the addition energy exceeded the single-particle level spacing[26]. In the p-channel away from the bandgap ($V_{gs}$~-30V), phase coherent transport and low contact barriers leading to Fabry-Perot like interference were observed[19] (Fig. S4). At ~50K, we observed conductance plateaus spaced by ~$e^2$/h in GNR1 and other ribbons (Fig. S10 and Supplementary Information), likely due to subbands in graphene nanoribbons as suggested previously[38,39].

Well defined quantum transport features were also observed in longer graphene nanoribbons ($L$>~100nm), although less frequently, suggesting higher likelihood of defects in longer nanoribbons. Fig. 3 shows a nanoribbon device (GNR2) with a longer $L$~140nm channel ($w$~17nm, Fig. 3a, inset), exhibiting $G$~$\frac{4e^2}{h}$ in the p-channel and peak field-effect mobility $\mu$~3200cm$^2$/Vs at room temperature. Variable temperature measurements again confirmed a single sharp dip in conductance near the bandgap (Fig. 3a) and $E_g$~$60\pm17$meV was estimated (Fig. S1). At T~3.3K, the conductance was suppressed near the bandgap, and regular Coulomb diamonds on both hole and electron branches were observed, separated by two relatively large diamonds similar to GNR1 (Fig. 3c). In the hole branch, we observed up to 7 regular diamonds with sizes of the diamonds or single electron addition energy $E_{add}$ following an even-odd pattern[22] (Fig. 3b). The even numbers of diamonds were larger than the corresponding odd diamonds because of the extra single-particle level spacings, which could be readily extracted (Table S2). In carbon nanotubes, electronic states are 4-fold degenerate because of spin and valley degeneracy, and four-fold shell filling have been observed[33,37]. In nanoribbons however, valley degeneracy is lifted due to different boundary conditions[1,2], resulting in two-fold spin degenerate states. The extracted energy level spacings (Table S2, Fig. 3f) agreed qualitatively with our theoretical calculations based on two-fold degenerate states in graphene nanoribbons.

We observed a wealth of well-defined excited states up to ~20meV in nearly all the Coulomb diamonds (Fig. 3d, f), and assigned them to the single-particle energy level spacings based on the ground state configuration of the quantum dot and our calculations (Supplementary Information). Using the same modeling approach as for GNR1, we found that the $\Delta\varepsilon_{n1}$ were again on the same order of magnitude with experimentally observed excite sates spectra as well as the size of even-odd diamonds (Table S2). We also carried out numerical simulation of Coulomb diamonds and excited states to quantitatively match our experiments (Fig. 3e & Supplementary Information). The origins of some excited states are unclear and require further investigation, such as the three lines terminated on diamond 0 with energies ~8meV, 18meV and -13meV, respectively (Fig. 3d). These states have much higher energy than $\Delta\varepsilon_{21}$ and are possibly due to interaction effects[22].

In Fig. 4, we show transport data for a third graphene nanoribbon device (GNR3, $w$~14nm, $L$~60nm, Fig. 4a inset), which also exhibited a high p-channel conductance and a sharp dip near the Dirac point at low temperatures with an estimated $E_g$~49$\pm$15meV (Fig. S2). At 4.2K, the differential conductance plot near the Dirac point showed a single large diamond corresponding to the bandgap, albeit with a gate switching event at $V_{gs}$~8V (Fig. 4b).

Our control experiments found that lithographically patterned graphene nanoribbons[30]

generally showed lower conductance and defect dominant transport characteristics at low temperatures (see Supplementary Information and Fig. S6), similar to previous reports[5-9]. A fraction of nanoribbon devices from unzipped nanotubes did not show well-defined quantum transport signatures, especially for long nanoribbons (Supplementary Information and Fig. S7, S8). These nanoribbons also exhibited lower conductance and mobility, likely due to lower ribbon quality.

Taken together, our results show that quantum transport features of graphene nanoribbons are highly reflective of the ribbon quality. We note that a recent paper reported improved quality of GNRs derived from heavily oxidized nanotubes by annealing[31]. However, signatures of transport gap were still present in those nanoribbons. The room temperature on-state conductivity of GNR1 and GNR2 shown here is ~700 and ~800 times higher than a typical device reported in Ref. 31. Our graphene nanoribbons do differ from carbon nanotubes with a fraction of ribbons exhibiting conductance exceeding $4e^2/h$ and two-fold electron shell filling, and from previous nanoribbons without overwhelming effects of the transport gap. High quality graphene nanoribbons are new types of quantum wires for exploring new physics (such as magnetic edge states[2,3]) and device concepts (such as spin qubits[4]) not present in seamless nanotubes.

**Methods**

**Graphene nanoribbon making**

We synthesized the high quality nanoribbons from multi-walled carbon nanotubes following Ref. 10. Briefly, multi-walled carbon nanotubes (Aldrich 406074-500MG, produced by arc discharge method, diameter: 4-15 nm, number of walls: 5-20) were calcined at 500°C for 2 h. We then dissolved the calcined nanotubes (15 mg) and 7.5 mg poly(m-phenylenevinylene-co-2,5-dioctoxy-p- phenylenevinylene) (PmPV) in 10ml 1,2-dichloroethane and sonicated for 1 h. The solution was ultracentrifuged at 40000 r.p.m. for 2 h and the supernatant was collected for experiments. Most nanoribbons in the final products were 1-2 layers[10,11].

**Graphene nanoribbon device fabrication**

We spun the high quality nanoribbon solution on 300nm $SiO_2/p^{++}$ Si substrate with pre-patterned metal markers, and used AFM to locate individual 1-2-layer nanoribbons. Although some nanotubes were also deposited on substrate, they were quite easy to be recognized due to much larger apparent height (>~4nm) than nanoribbons (<~1.8nm) under AFM (Fig. 1a, b). Extra care was taken to avoid nanotubes during device fabrication process, confirmed again by AFM on finished devices. We used electron beam lithography to pattern source/drain, evaporated 20nm Pd and did metal lift-off to form metal leads. The devices were finally annealed in Ar at 200°C for ~15mins to improve the contacts.

**Low temperature measurement setup**

The graphene nanoribbon devices were mounted in a variable temperature inset for low temperature measurements. We measured the G-$V_{gs}$ characteristics of the nanoribbon devices during cool downs by a standard semiconductor analyzer (Agilent 4156C) with low bias of $V_{ds}$=1mV. Below ~50K, we switched the measurement setup to a standard lock-in setup. We used two separate programmable DC sources (Keithley 237) as $V_{ds}$ and $V_{gs}$ and measured the differential conductance by a lock-in amplifier (Stanford Research Systems SR830).

**TEM**

TEM samples were made on porous silicon grids (SPI Supplies, US200-P15Q UltraSM 15nm Porous TEM Windows). TEM was performed with an FEI Tecnai G2 F20 X-TWIN. The operating voltage was 200kV.

## Acknowledgements

We thank David Goldhaber-Gordon for helpful discussions. The work at Stanford is

supported in part by Office of Naval Research (ONR), the ONR Graphene MURI, MARCO

MSD Focus Center and Intel. The work at University of Florida is supported in part by NSF and Office of Naval Research.

**Author contributions**

X. W. and H. D. conceived and designed the experiments. X. W. and J. W. fabricated the devices, performed the experiments and analyzed the data. Y. O. and J. G. performed simulations. L. J. provided graphene nanoribbon samples. H. W. and L. X. performed TEM characterizations. X. W., Y. O., J. G. and H. D. co-wrote the paper. All authors discussed the results and commented on the manuscript.

**Competing financial interests**

The authors declare that they have no competing financial interests.

**Additional Information**

Supplementary information accompanies this paper at

www.nature.com/naturenanotechnology. Reprints and permission information is available online at http://npg.nature.com/reprintsandpermissions/.

Correspondence and requests for materials should be addressed to H. D.

**FIGURE CAPTIONS**

**Figure 1.** High quality unzipping derived graphene nanoribbons. **(a)** AFM images of a typical high quality as-made graphene nanoribbon ($w$~27nm) next to a carbon nanotube on substrate. The obvious difference in height could be used to distinguish them. Our nanoribbons are typically ~0.3-0.6nm higher than those made from exfoliated graphene with the same number of layers due to PmPV coatings introduced in the synthesis[10,13]. H, height; W, width; d, diameter. The trace of the nanoribbon and nanotube appear about equally wide because (1) the trace is more parallel to the nanotube and (2) the AFM tip-size effect[15] depends on the height of the structure, higher nanotube causes more widening due to the conical shape of the AFM tips. **(b)** AFM image of GNR1 ($w$~14nm) discussed in the main text before device fabrication. We did careful AFM after device fabrication to ensure that only the nanoribbon was connected by the leads. **(c)** TEM image of a typical $w$~17nm high quality graphene nanoribbon with sub-nanometer edge roughness.

**Figure 2.** Electron transport of GNR1 ($L$~86nm). **(a)** Room temperature low bias ($V_{ds}$=1mV) $G$-$V_{gs}$ characteristics of GNR1. Lower inset shows the AFM image of the device. Upper inset shows $G$ vs. $T$ at $V_{gs}$=$V_{Dirac}$-35V in the hole channel. The metallic behavior, also observed in high quality carbon nanotube devices[20], suggests that the Pd contact is ohmic to the valence band, and the lower resistances at lower temperatures is due to reduced scattering by thermal depopulation of acoustic phonons. **(b)** Low bias ($V_{ds}$=1mV) $G$-$V_{gs}$ characteristics of GNR1 under various temperatures down to 60K. Inset shows zero-bias $G$-$V_{gs}$ characteristics at 2K. **(c)** Color scale differential conductance vs. $V_{ds}$ and $V_{gs}$ near the bandgap, showing single electron charging behavior. A gate switching was present near $V_{gs}$~1.5V indicated by the arrow. **(d)** Differential conductance in the electron branch near the bandgap, showing regular

Coulomb diamonds with excited states. The number of electrons in the quantum dot is marked for each diamond. The excited states energies $\Delta\varepsilon_{n0}$ (for the $n$th excited state relative to the ground state in unit of meV) are also marked. **(e)** Top panel: differential conductance in the hole branch near the bandgap. The number of holes in the quantum dot is marked for each diamond. Bottom panel: zero $V_{ds}$ line cut from the top panel, showing the peak pairing and enhanced conductance in the odd-numbered diamond valleys, a signature of Kondo effect. The spin configurations are also marked for each valley. Inset: constant $V_{gs}$ line cut in the middle of the third hole diamond in the top panel. The conductance is enhanced at zero bias as expected for Kondo effect[28].

**Figure 3.** Electron transport of a high quality quantum dot in GNR2 ($L\sim140$nm). **(a)** Low bias ($V_{ds}$=1mV) $G$-$V_{gs}$ characteristics under various temperatures down to 50K. Inset shows the AFM image of the device. **(b)** Experimentally measured single electron addition energy $E_{add}$ as a function of number of holes in the quantum dot, with an even-odd pattern. A small gate switching event happened in diamond 6, and $E_{add}(6)$ was measured after correcting the switching. Single-particle level spacings could be extracted. For example $\Delta\varepsilon_{10} = E_{add}(2) - E_{add}(1)$, $\Delta\varepsilon_{21} = E_{add}(4) - E_{add}(3)$. **(c)** Differential conductance as a function of $V_{gs}$ and $V_{ds}$ at 3.3K near the bandgap. The number of electrons and holes in the quantum dot are marked. **(d)** High resolution differential conductance scan across diamond 0 and 1, clearly showing excited states. The excited states are marked and assigned to the corresponding single-particle level spacings. **(e)** Simulated differential conductance of the same area in **(d)** at $T$=5K. See supplementary information for details. **(f)** Differential conductance scan for 6 Coulomb diamonds on the hole branch. The number of holes and ground state configuration for each diamond are illustrated. All the excited states are marked

and assigned to the corresponding energy level spacings. See Fig. S9 for the raw data without analysis blocking the features.

**Figure 4.** Electron transport of GNR3 ($L$~60nm). **(a)** Low bias ($V_{ds}$=1mV) $G$-$V_{gs}$ characteristics under various temperatures down to 50K. Inset shows the AFM image of the device. **(b)** Differential conductance as a function of $V_{gs}$ and $V_{ds}$ near the bandgap, showing single electron charging behavior. The central diamond (with size ~55meV as marked by the solid blue lines) corresponds to the bandgap of the nanoribbon free from any mid-gap states. There is a gate switching near $V_{gs}$~8V marked by the arrow.

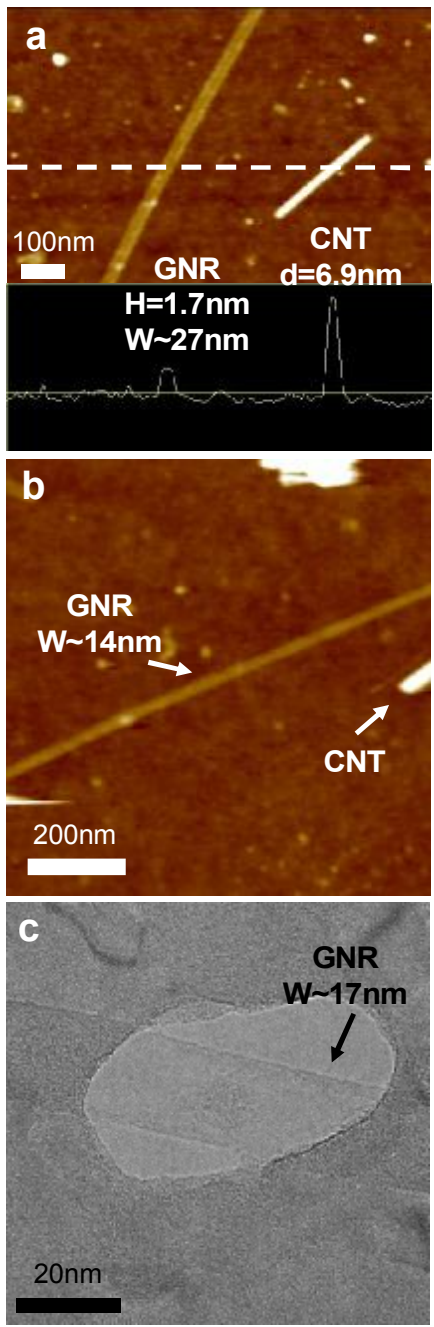

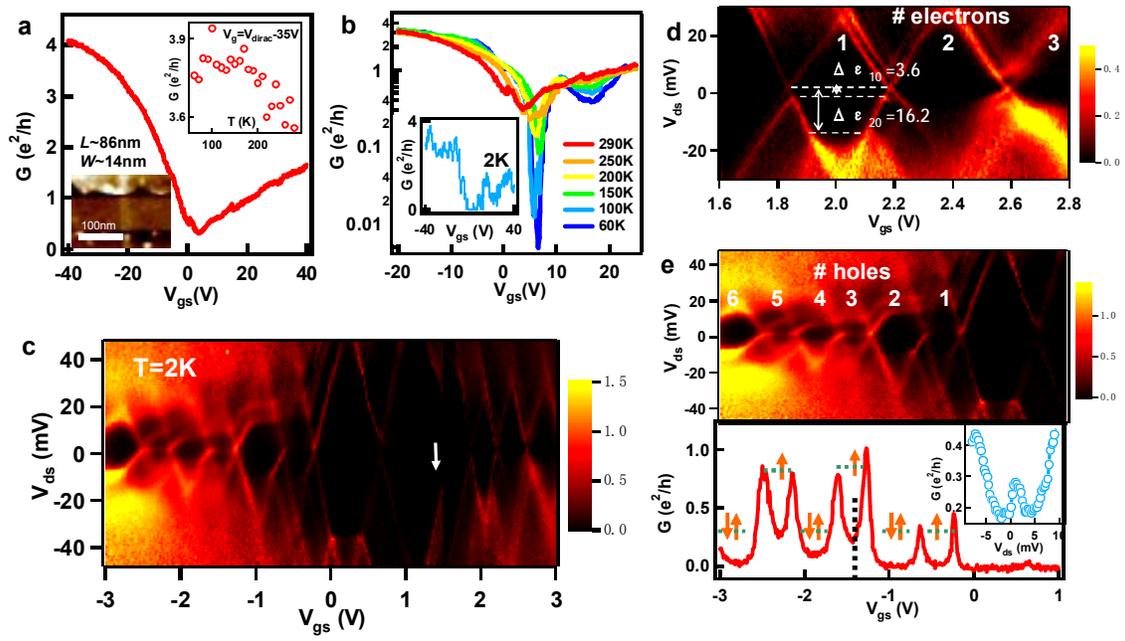

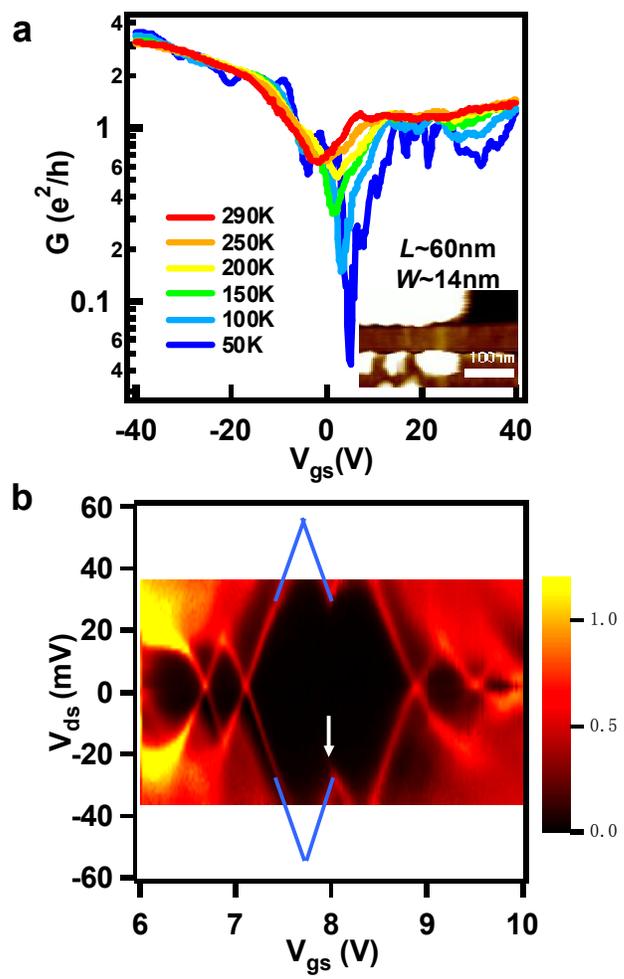

**a**

G (e²/h)

- 290K
- 250K
- 200K
- 150K
- 100K
- 50K

L~60nm
W~14nm

100nm

V_gs(V)

**b**

V_ds (mV)

V_gs (V)

# Graphene Nanoribbons with Smooth Edges Behave as Quantum Wires

Xinran Wang, Yijian Ouyang, Liying Jiao, Hailiang Wang, Liming Xie, Justin Wu, Jing Guo, and Hongjie Dai

Supplementary Information:

1. **High resolution aberration-corrected TEM image of a high quality GNR.**

2. **Extraction of $E_g$ for the high quality GNR devices using NEGF simulation.**

3. **Transport data of GNR1 ($L$~86nm) at 4.2K in a different cool down.**

4. **Phase coherent transport in GNR devices.**

5. **Low temperature conductance of wider GNR devices.**

6. **Calculation of the excited states energy and comparison with experiments.**

7. **Simulation of Coulomb blockade pattern of GNR2 and comparison with experiments.**

8. **Electron transport data of a lithographic GNR.**

9. **Additional electron transport data of unzipping derived GNRs.**

10. **Raw data in Figure 3 of the main text without the dashed lines.**

11. **Conductance plateaus of GNR1 and GNR2 at low temperatures.**

## 1. High resolution aberration-corrected TEM image of a high quality GNR.

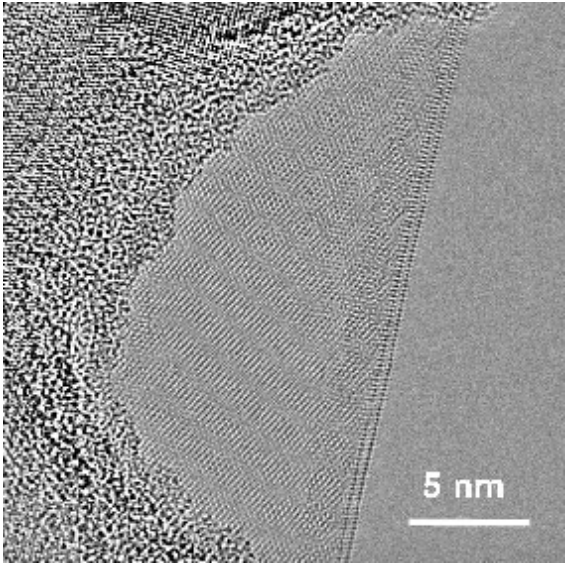

Fig. S1. Aberration-corrected TEM image of a typical bi-layer, non-AB stacked GNR in the sample used for this work with apparently very smooth edge. The data was taken under an operation voltage of 80 kV on TEAM 0.5 at the Lawrence Berkeley National Laboratory. For systematic TEM and Raman data of GNRs produced by the nanotube unzipping method (ref.10 of main text), see ref. 11 of the main text.

## 2. Extraction of $E_g$ for the high quality GNR devices using NEGF simulation.

The $G$-$V_{gs}$ characteristics in Fig.2a suggest asymmetrical SB heights for hole and electron transport. The SB height for holes is smaller than that for electrons, and is likely to be negative. In order to extract the bandgaps from the measured minimum conductance at different temperatures, the quantum transport equation is solved in the NEGF formalism with a self-consistent potential [1] to compute the minimal conductance, in which the self-consistent potential is obtained by a three-dimensional Poisson solver. The channel conductance is computed by the Landau formula, $G = 2e^2/h \int Tr(E)[-\partial f/\partial E]dE$ , where $Tr(E)$ is the transmission computed by the NEGF formalism, and $f(E) = 1/(1+\exp((E-E_F)/k_BT))$ is the Fermi-Dirac distribution function. The Hamiltonian is described by the Dirac Hamiltonian with quantized transverse wave vectors. The simulated ballistic conductance is multiplied by a gate-voltage- and temperature- independent transmission $0<T_r<1$, to fit the experiment, which models the effect of scattering in the GNR channel. This is a simplified treatment of scattering but is expected to have a negligible effect on the extracted value of $E_g$, since thermionic emission over and tunnelling through barriers play a dominant role in the extraction process as described below.

The procedure to extract $E_g$ is described as follows. The minimum conductance for a given bandgap is found for each temperature point (in the range of 70K to 290K) by the self-consistent NEGF simulation. A root mean square (RMS) error is defined as $\sigma = (<[(G(T) - G_{exp}(T))/G_{exp}(T)]^2>)^{1/2}$, where $G(T)$ and $G_{exp}(T)$ are the simulated and experimental conductance values at $T$, respectively, and the average is taken over different temperatures. We found that the slope of the $\log(G)$ vs. $-1/T$ curve is dominantly determined by the bandgap for negative SB height values, and it is independent of the exact value of $T_r$. After a group of curves of different bandgaps are simulated, the best fitting bandgap (i.e., the extracted bandgap) is given by the curve that has the smallest RMS error, $\sigma_{min.}$

For the three high quality GNR devices, the extracted bandgaps by assuming single-layer GNRs are $E_g = 72$ meV, $E_g = 60$ meV and $E_g = 49$ meV for GNR1, 2 and 3 respectively. By allowing a fitting error of $2\sigma_{min}$, the extracted $E_g$ of the three devices can be varied by $\pm 18$ meV, $\pm 17$ meV, and $\pm 15$ meV, respectively, from their best fitting values. If a twisted bilayer GNR is considered, the Hamiltonian of a twisted bilayer graphene can be simplified to an effective Dirac Hamiltonian similar to a single-layer based on the perturbation theory, but with a renormalized Fermi velocity, where the renormalization factor is about 0.8 [2]. By using a renormalization factor of 0.8, the extracted $E_g$ by the best fitting reduces by 10 meV and 5 meV for GNR1 and GNR2, respectively, which are within the error range indicated above.

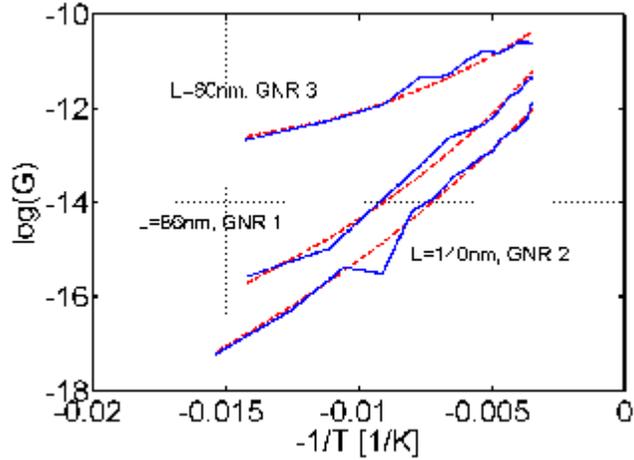

**Figure S2.** Fitting the measured minimum conductance (blue solid lines) in log scale as a function of $-1/T$ by the NEGF simulation (red dashed lines) for three high quality GNR devices. The temperature varies from 70K to 290K. The extracted bandgaps by the best fitting are $E_g$=72 meV, $E_g$=60 meV and $E_g$=49 meV for GNR1, 2 and 3 as defined in the main text respectively.

### 3. Transport data of GNR1 (*L*~86nm) at 4.2K in a different cool down.

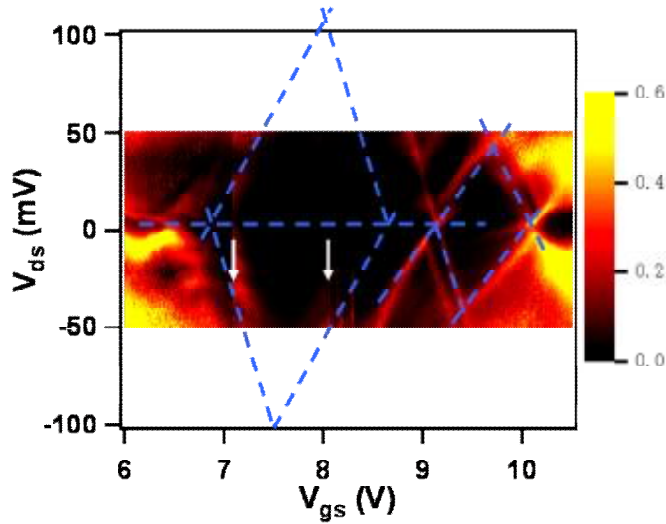

**Figure S3.** Differential conductance as a function of $V_{gs}$ and $V_{ds}$ near the bandgap of GNR1 in the main text, taken in a separate cool down at 4.2 K. Only one central diamond appeared in the data. There were two gate switching events marked by the white arrows. The blue dashed lines were drawn as a guide to the eye for the big diamond (left) corresponding to the bandgap and a small diamond (right).

## 4. Phase coherent transport in GNR devices.

We observed phase coherent transport in several GNR devices in the p-channel away from the bandgap region. Figure S3 showed two such examples taken on GNR1 and GNR2. Differential conductance as a function of bias voltage $V_{ds}$ and gate voltage $V_{gs}$ showed Fabry-Perot like interference [3]. As marked by the white arrows, the characteristic energy scale of GNR1 and GNR2 is $V_c$~19 mV and 10 meV, respectively, in good agreement with the calculated value of $V_c = \dfrac{h v_F^0}{2L \cdot e} \sim 21\, mV$ and 13 meV, where $v_F^0 \sim 8.7 \times 10^5\, m/s$ is the Fermi velocity of 2D graphene and $L$ is the channel length [3].

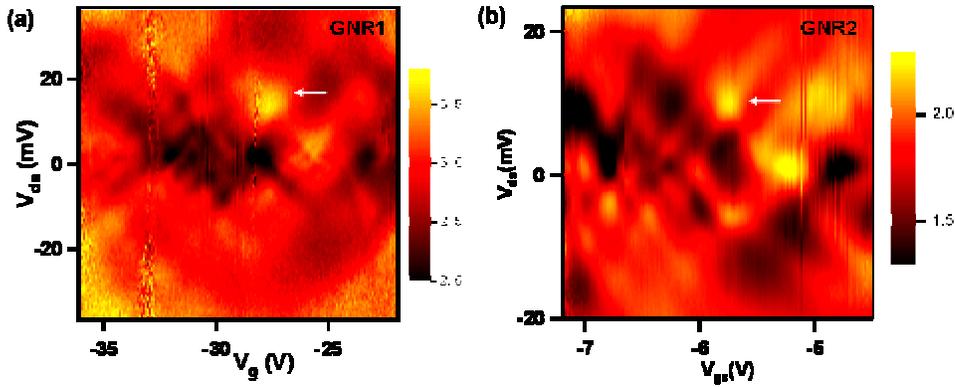

Figure S4. Differential conductance as a function of $V_{gs}$ and $V_{ds}$ in the hole channel away from the bandgap region, taken on **(a)** GNR1 and **(b)** GNR2, showing Fabry-Perot interference patterns. The white arrows point to the characteristic energy $V_c$ for both devices.

## 5. Low temperature conductance of wider GNR devices.

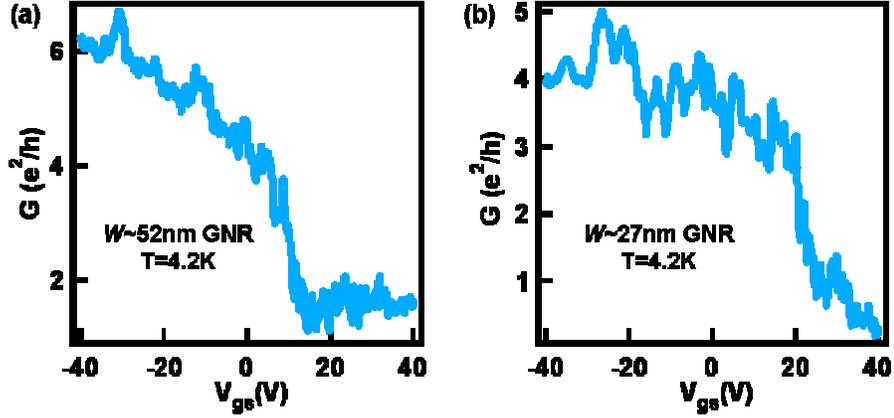

Figure S5. Zero bias differential conductance vs $V_{gs}$ for a w~52 nm **(a)** and w~27 nm **(b)** GNR devices at 4.2 K. At negative $V_{gs}$, they both exhibited higher conductance than $4e^2$/h, suggesting that more than two doubly degenerate subbands were involved in the transport. Wider GNRs tent to have more subbands involved probably due to the small subband spacings.

## 6. Calculation of the excited states energy and comparison with experiments.

Both single-layer and non-AB-stacked bilayer GNRs were examined. We used an atomistic $P_z$ orbital tight binding (TB) model with nearest neighbour interaction to compute the single-layer graphene *E-k* [1]. The bandgap and excited states of single-layer GNRs were derived by quantizing the *E-k* with particle-in-a-box boundary conditions, i.e. $k_{x,m}=m\pi/W$, and $k_{y,n}=n\pi/L$, where $k_x$ ($k_y$) is the wave vector along width (length) direction with regard to the Γ point of the graphene Brillouin zone, $W(L)$ is the GNR width (length), and *m and n* are positive integers. The QD energy levels derived from the lowest semiconducting subband can be approximated by $\varepsilon(k_{\parallel}(n)) = \pm\sqrt{(\hbar v_F \pi / 3W)^2 + (\hbar v_F n\pi / L)^2}$. For non-AB-stacked bilayer GNRs, we computed the bandgap and excited states by quantizing the wave vectors of non-AB-stacked bilayer graphene with Moiré patterns. The band structures of the non-AB-stacked bilayer graphene were computed using the TB parameters as described in Ref. [4]

Due to the uncertainties in the edge atomistic structures, number of layers, and layer stacking structure and simplicity of the calculation approach, the comparison to the experiment is qualitative and only carries an order of magnitude meaning. We did calculations for single-layer GNRs and bilayer GNRs with different rotation angles of Moiré patterns. Note that the set of the rotation angles is infinite and therefore only a few representative calculations are listed in Table S1 for GNR1 and Table S2 for GNR2. In the low resolution scan as shown in Fig. 3f, we were able to assign all the observed excited states to the expected level spacings (Table S2). We note that in the scan as shown in Fig. 3f, not all the expected levels (such as $\Delta\varepsilon_{21}$) show up probably due to low resolution.

Table S1: Calculated bandgap and excited states for specific structures compared to the experiment data of GNR1

| Rotation angle of bilayer Moiré pattern | GNR Band gap (meV) | $\Delta\varepsilon_{21}$(meV) | $\Delta\varepsilon_{31}$ (meV) |
|---|---|---|---|
| 32.2 | 55.3 | 6.6 | 18.3 |
| 21.8 | 61.5 | 2.0 | 13.5 |
| 13.2 | 70.6 | 7.9 | 13.3 |
| 9.43 | 54.8 | 10.6 | 14.8 |
| 6.01 | 58.2 | 11.5 | 14.3 |
| Single-layer GNR | 84.9 | 12.2 | 28.3 |
| Experiment | 72±18 | 3.6 | 16.2 |

Table S2: Calculated bandgap and excited states for specific structures compared to the experiment data of GNR2

| Rotation angle of bilayer Moiré pattern | GNR Band gap (meV) | $\Delta\varepsilon_{21}$ (meV) | $\Delta\varepsilon_{31}$ (meV) | $\Delta\varepsilon_{41}$ (meV) | $\Delta\varepsilon_{32}$ (meV) | $\Delta\varepsilon_{42}$ (meV) | $\Delta\varepsilon_{43}$ (meV) | $\Delta\varepsilon_{53}$ (meV) |
|---|---|---|---|---|---|---|---|---|
| 32.2 | 53.4 | 3.4 | 10.6 | 20.2 | 7.2 | 16.8 | 9.5 | 20.3 |
| 21.8 | 64.6 | 4.6 | 8.4 | 13.2 | 3.8 | 8.6 | 4.8 | 11.1 |
| 13.2 | 57.1 | 3.3 | 9.8 | 18.5 | 6.5 | 15.2 | 8.7 | 18.6 |
| 9.43 | 54.9 | 2.8 | 7.9 | 15.7 | 5.4 | 12.9 | 7.5 | 16.2 |
| 6.01 | 53.8 | 4.5 | 11.4 | 19.6 | 6.9 | 15.1 | 8.2 | 17.2 |
| Single-layer GNR | 70.1 | 6.1 | 14.7 | 24.8 | 8.6 | 18.7 | 10.1 | 18.9 |
| Experiment (from excited | 60±17 | 2.5-3.6 | 8.2-10.1 | 19.2-19.8 | 5.3-6.2 | 11.5-16.3 | 9.6-12.5 | 19.7 |
| state, Fig. 3d, f) | | | | | | | | |
| Experiment (from even-odd pattern, Fig. 3b) | n/a | 1.7 | 6.2 | 17.3 | 4.5 | 15.6 | 11.0 | n/a |

## 7. Simulation of Coulomb blockade pattern of GNR2 and comparison with experiments.

A single-electron charging simulator is developed based on the many-particle Fock space master equation [1]. The coupling of the QD to the source and drain causes the transition of the QD state from one many-body state to another, which can result in the source-drain current. The input parameters of the simulator are the single particle energy levels, coupling capacitances, and source/drain contact broadening. The output is the conductance (or current) as a function of the applied voltages. At a low temperature ($T$<5 K) and small magnitude of gate biases, the condition to use the master equation, $U_0$>> $k_B T$ and $\Gamma$, is satisfied, where $U_0$ is the single electron charging energy, and $\Gamma$ is the total broadening by the electrodes. The simulation captures Coulomb diamond shapes and sizes, as well as the excited state lines. By comparing the simulation results with experiments, one can clearly identify where each excited state line comes from as labelled in Fig. 3e. These excited state lines can stem from adjacent energy levels (e.g. $\varepsilon_2$-$\varepsilon_1$) as well as non-adjacent energy levels (e.g. $\varepsilon_3$-$\varepsilon_1$). One can also extract capacitances and single particle levels by the best fitting of the simulated results to the experimental measurements.

For GNR2, the gate capacitance is estimated as $C_g$~0.87 aF from the gate voltage periods of the conductance peaks at low $V_{ds}$. The drain capacitance $C_d$~2.13 aF is found through the slope of the Coulomb diamond boundaries of the negative slope sides. From the single electron charging energy, the total capacitance of the QD can be computed, and the source capacitance of $C_s$~2.70 aF is obtained. From the experimental data in Fig. 3d and simulated CB pattern in Fig.3e (with $C_g$=0.87 aF, $C_s$=2.70 aF $C_d$=2.13 aF, and a source/drain broadening of 0.05 meV), we can obtain the following energy spacing values, $\Delta\varepsilon_{41}$ ~19.4 or 19.8 meV, $\Delta\varepsilon_{31}$ ~10.1 meV and $\Delta\varepsilon_{21}$ ~2.5 or 3.6 meV.

## 8. Electron transport data of a lithographic GNR.

In comparison with high quality GNR devices from unzipping MWNTs, we lithographically patterned GNR devices from pristine single-layer exfoliated graphene with similar dimensions. At low temperatures, the lithographic GNRs typically showed defect dominant behavior distinct from our high quality GNRs. Fig. S5 presents the transport data of a representative lithographic GNR device.

Different from the high quality GNRs, lithographic GNR devices usually showed lower conductivity and mobility as shown in Fig. S5a ($\sigma_{on} \sim 0.19\,mS$, $\mu \sim 210\,cm^2/Vs$). The on state conductance upon cooling was constant down to $T^* \sim 100$ K, followed by rapid decrease at lower temperature (Fig. S5a, inset). Near the Dirac point, the lithographic GNR showed suppressed conductance over a relatively wide region in the $G$-$V_{gs}$ characteristics at low temperature ($\Delta V_{gs} \sim 10\,V$, ~10 times that of GNR1), with some resonances inside (Fig. S5b). Similar behavior has been observed in lithographic GNRs and graphene nanoconstrictions and attributed to transport gap resulting from edge disorders or charged impurities [5-9]. Theoretical calculations have shown that the density of states of GNRs with edge disorders are dominated by localized states near the Dirac point, and the charge transport could be through variable range hopping between these localized states at low temperature [10-12]. The sharp resonances in the transport gap are signatures of resonant tunneling through the localized states [5]. We note that the resonances in the transport gap only start to become obvious below the same $T^* \sim 100$ K (Fig. S5b). This is expected because when $T<T^*$, the charge carriers tend to localize near the defects in the aforementioned transport gap picture [5, 10], leading to the decrease in conductance. $k_B T^* \sim 8$ meV is also in good agreement with the characteristic temperature $T^*$ in Ref. 6. When $T>T^*$, $G_{min}$ appeared to follow a thermally activated behavior [5]. At 4.2 K, the transport features near the Dirac point were also highly different from the high quality GNRs (Fig. S5c). Over ~30 V span in $V_{gs}$ (the transport gap), the transport features were dominated by CB diamonds with sizes ranging from ~10 meV to ~30 meV. Some of the diamonds were not closed indicating multiple QD behavior [6, 7]. These features were similar to previously reported short channel lithographic GNRs [6-9]. In this case, the transport gap was dominant due to large numbers of localized states most likely by edge defects [6-12]. Coulomb diamonds did not close due to multiple dots in series as a result of defects along the GNR [6, 7]. Excited states were rarely observed in lithographically derived GNR QDs.

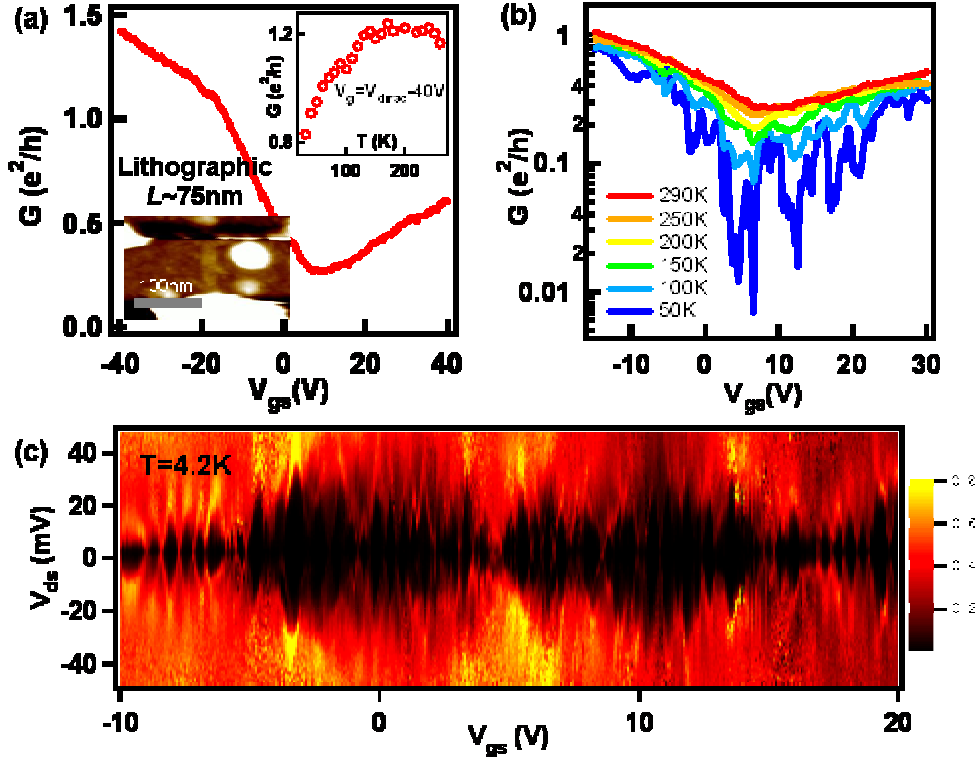

**Figure S6.** Transport measurement of a representative lithographic GNR device ($L$~75 nm). **(a)** Room temperature low bias ($V_{ds}$=1 mV) $G$-$V_{gs}$ characteristics of the lithographic GNR device. Lower inset is AFM images of the devices. Upper inset shows $G$ vs. $T$ at $V_{gs}$=$V_{Dirac}$-40 V in the p-channel. **(b)** Low bias ($V_{ds}$=1 mV) $G$-$V_{gs}$ characteristics of the lithographic GNR device at various temperatures down to 50 K. **(c)** Differential conductance of the lithographic GNR as a function of $V_{gs}$ and $V_{ds}$ in the transport gap at 4.2 K. In the transport gap, CB diamonds with size ranging from ~10 meV to ~30 meV are observed over ~30 V $V_{gs}$ span.

## 9. Additional electron transport data of unzipping derived GNRs.

Fig. S6 and S7 show electron transport data of two lower quality unzipping derived GNRs with $L$≥~175nm. The p-channel conductance was lower than ~2 $e^2/h$ for both devices at room temperature and decreased at lower temperatures, similar to lithographic GNRs. We measured the gap region near the Dirac point at $T$=4.2 K, and observed many small diamonds (some of them were irregular without complete closure) over a large $V_{gs}$ range, indicating the deviation from a single quantum dot behaviour likely due to defects on the ribbon.

Note that we have also measured some wider GNRs with $w$~20-30 nm. Down to the base temperature of our cryostat (~2 K), the conductance of these GNRs near the Dirac point was usually not depleted without Coulomb blockade likely due to the more metallic nature of these wider GNRs. This was also reflected from the weaker $G_{min}$ (conductance at the Dirac point) vs. -1/$T$ dependence than that of narrower ribbons in Fig.S1.

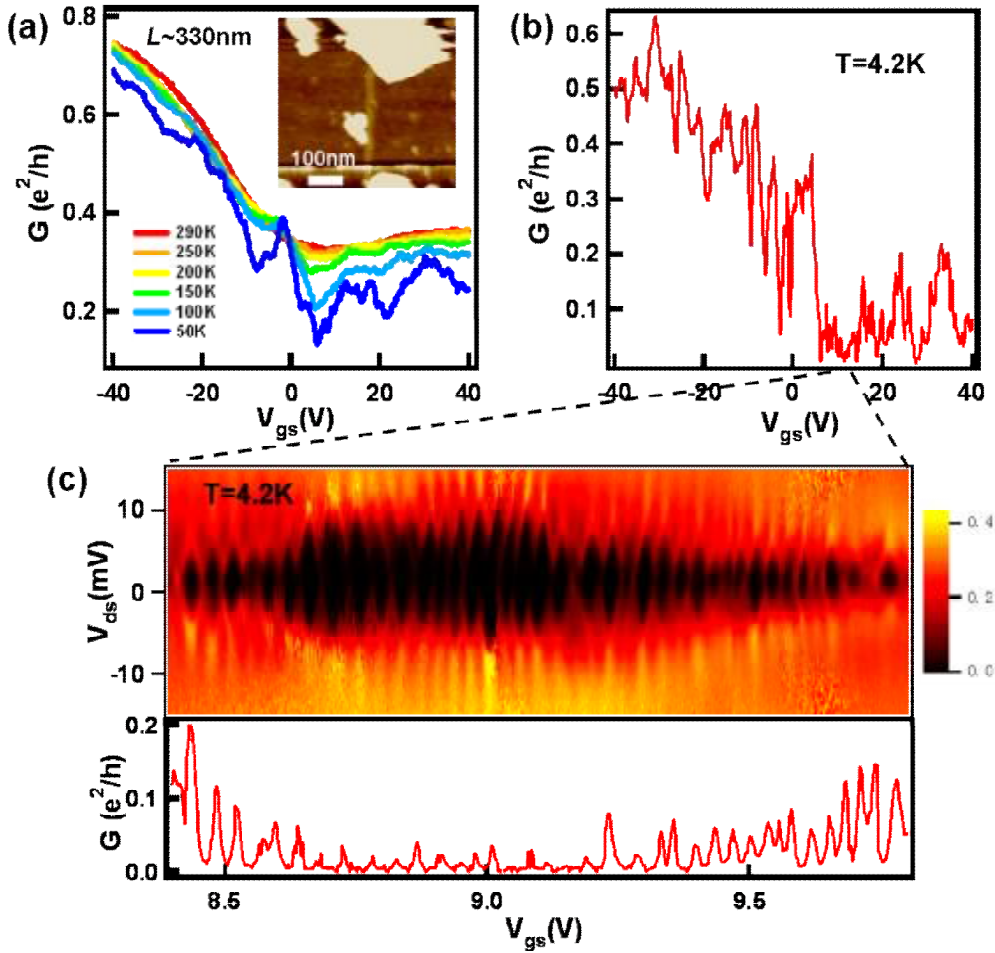

Figure S7. Electron transport data of a $w$~17nm, $L$~330nm unzipping derived GNR device. (a) Low bias ($V_{ds}$=1 mV) $G$-$V_{gs}$ characteristics of the GNR device under various temperatures

down to 50 K. Inset shows the AFM image of the device. (**b**) Zero bias differential conductance as a function of $V_{gs}$ of the GNR device at 4.2 K. (**c**) Top panel: differential conductance as a function of $V_{gs}$ and $V_{ds}$ of the GNR device near the Dirac point, showing many small diamonds without a clean, large diamond corresponding to the bandgap. Bottom panel: zero $V_{ds}$ line cut from the top panel.

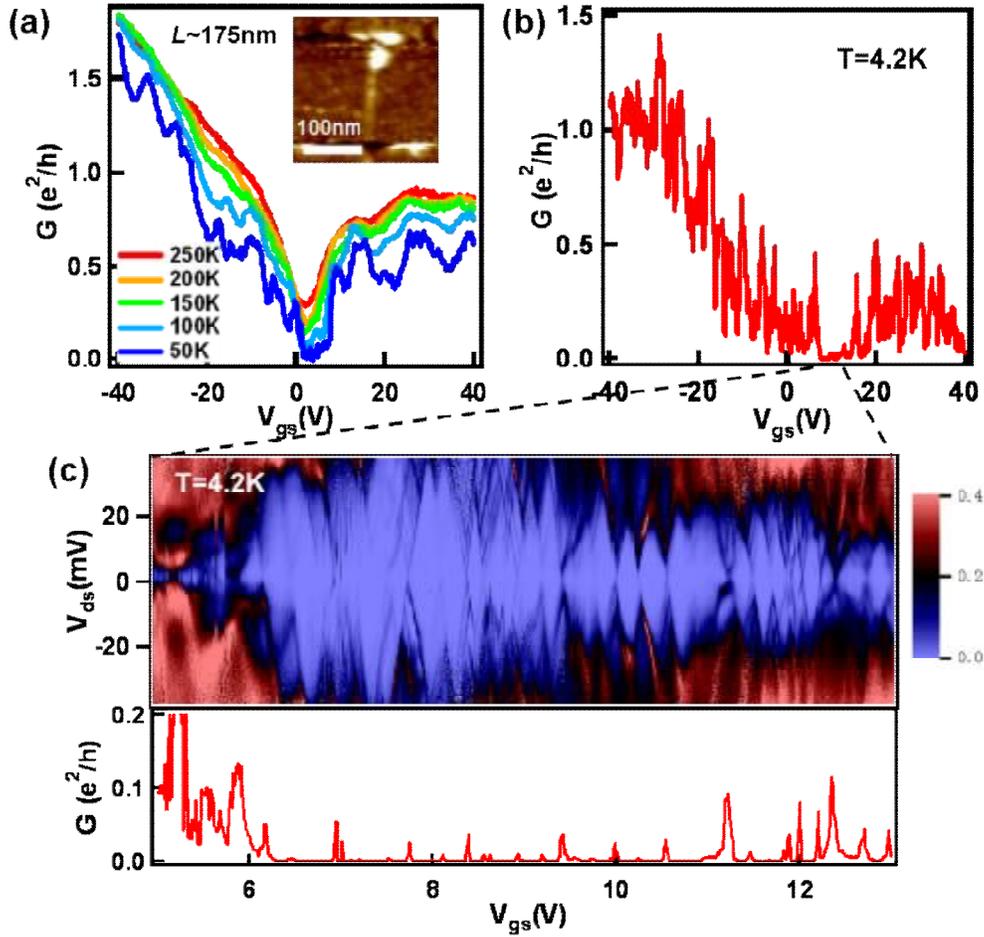

Figure S8. Electron transport data of a $w$~12 nm, $L$~175 nm unzipping derived GNR device. (**a**) Low bias ($V_{ds}$=1 mV) $G$-$V_{gs}$ characteristics of the GNR device under various temperatures down to 50 K. Inset shows the AFM image of the device. (**b**) Zero bias differential conductance as a function of $V_{gs}$ of the GNR device at 4.2 K. (**c**) Top panel: differential conductance as a function of $V_{gs}$ and $V_{ds}$ of the GNR device near the Dirac point, showing many small, irregular diamonds without a clean, large diamond corresponding to the bandgap. Bottom panel: zero $V_{ds}$ line cut from the top panel.

**10. Raw data in Figure 3 of the main text without the dashed lines.**

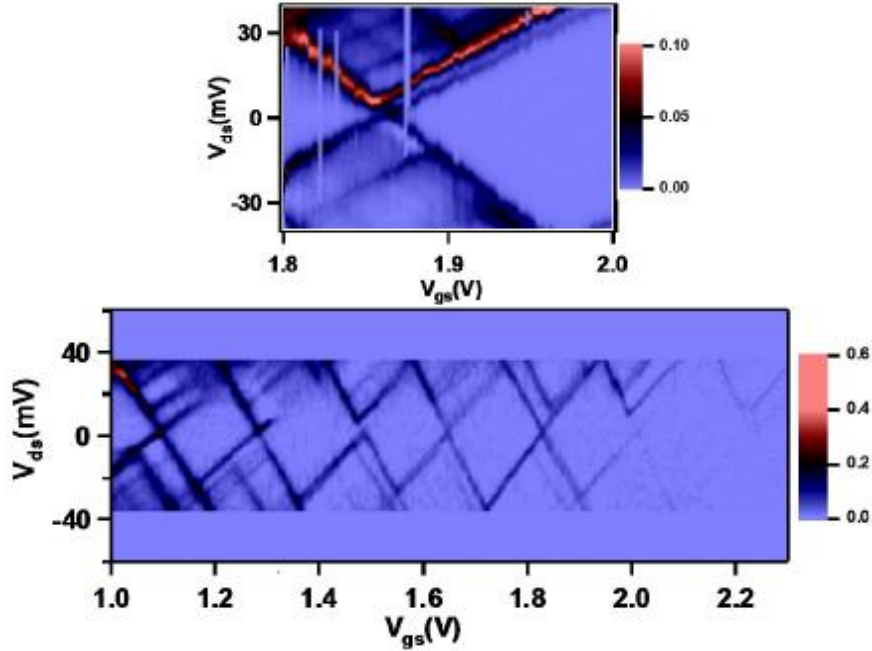

Figure S9. Differential conductance scans for GNR2 at 3.3 K without dashed lines. The data are the same as Fig. 3d and f in the main text.

**11. Conductance plateaus of GNR1 and GNR2 at low temperatures**.

We have observed possible signatures of multiple 1D subbands as conductance plateaus [13, 14] in several GNRs including GNR1 and GNR2 below ~100 K. The conductance plateaus were most clear at ~50-60 K, below which Fabry-Perot like oscillations started to kick in. Fig. S10 are $G$ vs $V_{gs}$ curves for GNR1 and GNR2 discussed in the manuscript. The conductance steps were ~0.9-1 $e^2$/h for both cases, which was much greater than previously observed step sizes in GNRs [13, 14]. Due to the two-fold degeneracy of subbands in GNRs, these data suggest that the transmission coefficient at each contact is ~0.7, partly due to reflection at the contacts, and up to 4 subbands are involved in the transport of GNR1 and GNR2. We note that transport through multiple subbands was rarely observed in carbon nanotubes. For wider ribbons, more subbands were populated based on the high conductance (up to ~7 e²/h) (Fig. S5), likely due to lower Schottky barriers to higher subbands.

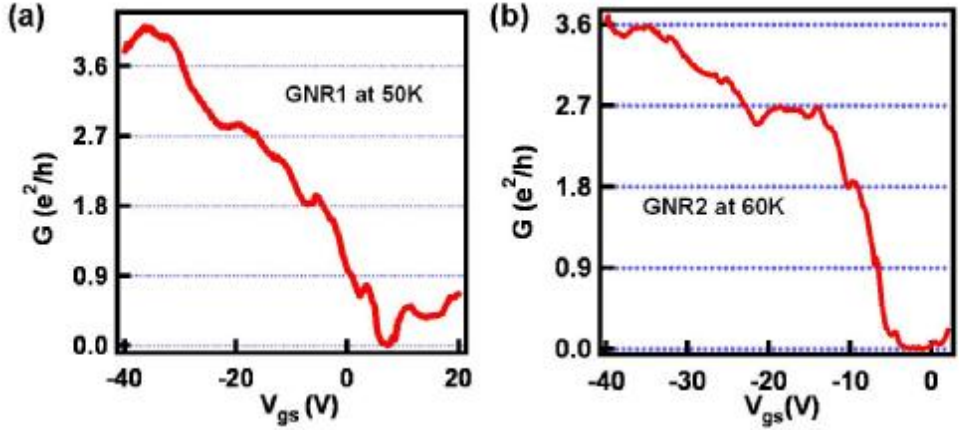

Figure S10. *G* vs *V_gs* curves for GNR1 at 50 K (a) and GNR2 at 60 K (b). Conductance plateaus were observed in both cases with steps of ~0.9-1 e²/h.